\documentclass{emulateapj-rtx4}

\begin{document}

\title{Insight-HXMT observations of 4U~1636-536: Corona cooling revealed with single short type-I X-ray burst}

\author{Y. P. Chen$^{1}$,  S. Zhang$^{1}$, S. N. Zhang$^{1,2}$, L. Ji$^{5}$, L. D. Kong$^{1}$,
X. L. Cao$^{1}$, Z. Chang$^{1}$, G. Chen$^{1}$,L. Chen$^{4}$, T. X. Chen$^{1}$
% ...
         }

\affil{$^{1}$ Key Laboratory for Particle Astrophysics, Institute of High Energy Physics, Chinese Academy of Sciences, 19B Yuquan Road, Beijing 100049, China}
\affil{$^{2}$ University of Chinese Academy of Sciences, Chinese Academy of Sciences, Beijing 100049, China}
\affil{$^{3}$ Department of Physics, Tsinghua University, Beijing 100084, China}
\affil{$^{4}$ Department of Astronomy, Beijing Normal University, Beijing 100088, China}
\affil{$^{5}$ Institut f\"{u}r Astronomie und Astrophysik, Kepler Center for Astro and Particle Physics, Eberhard Karls Universit\"{a}t, Sand 1, D-72076 T\"{u}bingen, Germany}

\email{chenyp@ihep.ac.cn, szhang@ihep.ac.cn}


\begin{abstract}
Corona cooling was detected previously from stacking a series of short type-I bursts occurred during the low/had state of atoll outburst. Type-I bursts are hence regarded as sharp probe to our better understanding on the basic property of the corona. The launch of the first Chinese X-ray satellite Insight-HXMT has large detection area at hard X-rays which provide almost unique chance to move further in this research field. We report the first detection of the corona cooling by Insight-HXMT from single short type-I burst showing up during {\bf flare} of 4U 1636-536.
This  type-I X-ray burst has a duration of $\sim$13 seconds and hard X-ray shortage is detected with significance 6.2~$\sigma$ in 40-70 keV.
A cross-correlation analysis between the lightcurves of  soft and hard X-ray band, shows that the corona shortage lag the burst emission by 1.6 $\pm$1.2~s.
These results are consistent with those derived previously from stacking a large amount of bursts detected by RXTE/PCA within a series of {\bf flares} of  4U 1636-536. Moreover,  the broad bandwidth of Insight-HXMT allows as well for the first time to infer the burst influence upon the continuum spectrum via performing the spectral fitting of the burst, which ends up with the finding that hard X-ray shortage appears at around 40 keV in the continuum spectrum.  These results suggest that the evolution of the corona along with the outburst{\bf /flare} of NS XRB may be traced via looking into a series of embedded type-I bursts by using Insight-HXMT.
\end{abstract}
\keywords{stars: coronae ---
stars: neutron --- X-rays: individual(4U~1636-536) --- X-rays: binaries --- X-rays: bursts}

\section{Introduction}

Type-I X-ray burst, also named thermonuclear burst (hereafter burst),
is caused by unstable burning of the accreted hydrogen/helium on the
surface of a neutron star (NS) enclosed in an X-ray binary (XRB), and manifest themselves as a sudden increase (typically by a factor of 10 or greater) in the
X-ray luminosity followed by an exponential decay (for reviews, see \citealp{Lewin,Cumming,Strohmayer,Galloway}).
The most luminous bursts are the photospheric radius expansion (PRE) events,
for which the peak flux is comparable
to the Eddington luminosity at the surface of the NS.

More and more observational examples and theory models on the
burst influence on the persistent/accretion emission
have rapidly accumulated over recent years \citep{Degenaar2018}.
 Basically, there are three observational evidence and corresponding physical processes initiation between the two kinds emission:
 accretion rate change because of Poynting-Robertson drag induced by dynamic/light-pressure of burst;
 disk or/and corona change both in structural or/and intrinsic characteries owing to the burst cooling/heating;
 enhancement emission result from refection of the accretion disk.
Among the observation evidences, most of them are from inter-mediate long burst/super-burst or stacking lots of normal burst \citep{Degenaar2018}.

For the interaction between bursts and corona, the shortage in the hard X-ray  of the continuum emission is reported on several XRB, i.e., IGR~J17473-2721 \citep{chen2011,chen2012}, Aql~X-1 \citep{chen2013,maccarone2003}, 4U~1636-536 \citep{ji2013},  GS~1826-238  \citep{ji2014a},  KS~1731-260 \citep{ji2014b}, 4U~1705-44 \citep{ji2014b} and 4U 1728-34 \citep{Kajava2017}, based on RXTE/PCA and INTEGRAL observation.
It is very hard to detect the deficit up to 5$\sigma$ in
single short burst because of the relatively small detection area of the previously missions at hard X-ray, and hence
these firm detection significance are aggrandized by stacked tens to hundreds bursts.
For example, 4U~1636-536, the deficit detection is based on {\bf 36} bursts results.

In the timing analysis zone, effects of X-ray bursts on kHz QPOs are also revealed in several bursters such as Aql X-1 \citep{yu1999} and 4U~1636-536 \citep{peille2014}, often manifests as QPO frequency and persistent flux changes between before and after the burst. Such detections are interpreted
that the burst blew away or pulled up the inner disk based on the different viscous time-scale for QPO recovery.

4U~1636-536, a low-Mass X-ray Binary (LMXB), discovered with the 8th Orbiting Solar Observatory (OSO-8; \citealp{Swank1976}),  it is a well-studied LMXB which holds an 18th magnitude blue
star, V801 Ara in an orbit of 3.8 hr
\citep{van1990}.
It is one of the few persistent X-ray sources in our Galaxy that undergoes  
{\bf regularly transitions between the hard state and the soft state},
in a repeating period of roughly 70 days, and many bursts accompanied.
The properties of the burst oscillations of 579.3 Hz and super-bursts with hours duration have been detected and analyzed (see \citealp{Galloway} for a review).
From its colour-color diagram
(CCD), 4U~1636-536 traces a U-shape
or C-shape as a typical atoll source (detailed CCD was
shown in \citealp{Zhang2011}).
Its distance was estimated as $\sim$6~kpc \citep{Galloway2006} by using the PRE burst.

In this work we study the burst-corona interaction in the LMXB
4U~1636-536 using the first year of data collected with Insight-HXMT \citealp{Zhang2014}. In Sect. 2 we present
Insight/HXMT observations and data analysis procedure by detail.
In Sect. 3,  the shortage in hard X-ray band are given based on the broad-band spectrometry results.
In Sect. 4,  we present our interpretation  on the detection and the comparison with previous detection on the hard X-ray shortage during burst.

\section{Observations and Data analysis}

On June 15th of 2017, the Hard X-ray Modulation Telescope (HXMT, also
dubbed as Insight-HXMT,\citealp{Zhang2014})was launched in Jiuquan Satellite Launch Center. It excels in broad energy band (1-250 keV) detection ability and large effective area in hard X-rays energy band.
It consists of three
slat-collimated instruments: the High Energy X-ray
Telescope (HE), the Medium Energy X-ray Telescope
(ME), and the Low Energy X-ray Telescope (LE), with  collecting-area/energy-range  in~$\sim$5000~$cm^2$~in 20-250 keV, $\sim$900 $cm^2$~in 5-30 keV and $\sim$400~$cm^2$~in 1-10 keV respectively.

In this work,  we analyze the brightest of three bursts, represent the time-resolved spectroscopy and give our interpretation on the uniqueness observational hehavior.
By virtue of quick read-out time of Insight-HXMT detectors, there is little pile-up event at the PRE burst peak.
HEASOFT version 6.22.1 and Insight-HXMT Data Analysis software
(HXMTDAS) v2.01 were used to process and
analyze the data.
Only the small field of view (FoV) of LE and ME were used, because large FoV
were easily contaminated by near-by source and the bright earth.
The good time interval were filterd
 with the following criteria: (1) pointing offset angles
$<$0.05 degree; (2) elevation angles$>$6 degree; (3) the value
of the geomagnetic cutoff rigidity $>$ 6.

Among its first years observations, the total Insight-HXMT observation of 4U~1636-536 is $\sim$370 ks, covering the time span between February 11th and July 1st of 2018.
Eight type-I X-ray bursts\ref{table}
 are detected in 4U~1636-536.
  Among them , the first three burst have low flux level of persistent both in soft at $\sim$ 40 mCrab and hard X-ray at $\sim$ 15 mCrab.
 The other five bursts have high hard X-ray flux and relatively low flux in soft X-ray, indicates it locates island state (similar with low/hard state of black hole XRB).
Most of them are lack of LE results because of optical pollution from the bright earth.
 Form these eight bursts, we choose a burst which satisfies the  good-time-interval selection criteria of LE, ME and HE simultaneously,
 and locates at the island state  with a high hard X-ray flux $\sim$ 75 mCrab of its inhabited persistent emission. (Fig.\ref{fig_outburst_lc})
 The obsid number is P011465402801-20180701-01-01, with peak flux happened at MJD 58300.717896.



The lightcurve profile of the burst is derived in time
bins of 1 s in the full passband of LE\&ME and 40-70 keV of HE, with pre-burst emission subtracted, is shown in Fig \ref{fig_lc}, the top, middle and bottom pad for LE, ME and HE respectively.
From Fig. \ref{fig_lc}, the lightcurves are stable before and after the burst, indicating little variation of the persistent and background emission.

\begin{figure}[t]
\centering
   \includegraphics[angle=0, scale=0.4] {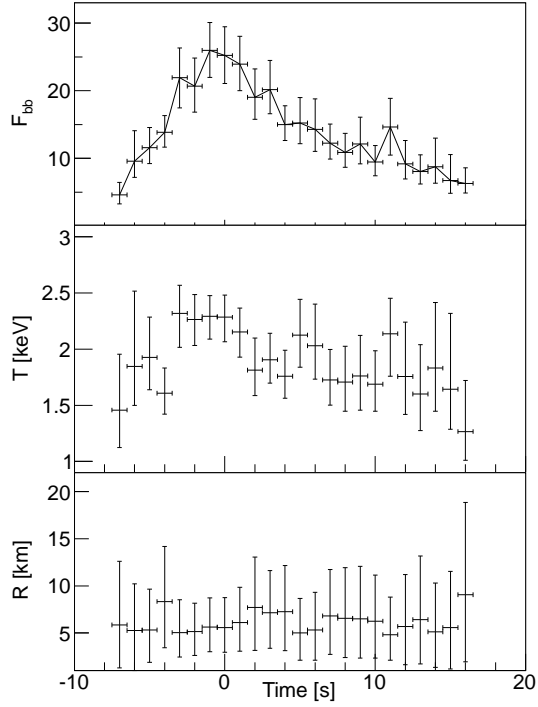}
 \caption{ The time evolution of blackbody bolometric flux, the temperature T, the observed radius of NS surface at 6 kpc,  with a time resolution of 1 second, the bolometric flux of blackbody $F_{bb}$ is in unit of $10^{-9}~erg/cm^{2}/s$.
  }
\label{fig_fit}
\end{figure}

We adopted the standard analysis procedure of burst, i.e., take pre-burst emission (including instrumental background and persistent/accretion flux of the neutron star system) as background to investigate the burst spectra evolution.
We divide the burst into intervals
of 1 seconds after the burst onset, and  extract the spectra of LE, ME and HE respectively.
For the burst, we use the time of the bolometric flux peak as a reference (0 second in Fig.\ref{fig_fit}) to produce the lightcurve/spectra.
A blackbody model (bbodyrad in Xspec) with fixed absorption 0.41$\times10^{22}~cm^{-2}$ as derived in \citet{Agrawal2016} is used to fit the burst spectra.
To compromise the effective area calibration deviation, a constant is added to the model.
At first attempt, for LE, the constant is fixed to 1, the others (ME and HE) are alterable during spectra fitting. But the fitting result indicate that most of the constant of HE are not convergent, owing to the low-significance of the HE detection. Under this situation, the constant of HE was fixed at 1 for the combined-spectra fitting, which based on combined-spectra fitting of Crab observations by the same detector selection.
The unabsorbed
bolometric flux of the spectra is estimated by the XSPEC model cflux,
and the observed blackbody radius is estimated under condition that the distance of 4U~1636-536 is 6 kpc.
The fitting results are shown in Fig \ref{fig_fit} and Fig \ref{spec}.


\section{Results}


\subsection{Lightcurves of burst emission in soft and hard X-ray band}

\begin{figure}[t]
\centering
      \includegraphics[origin=c,angle=0, scale=0.41]{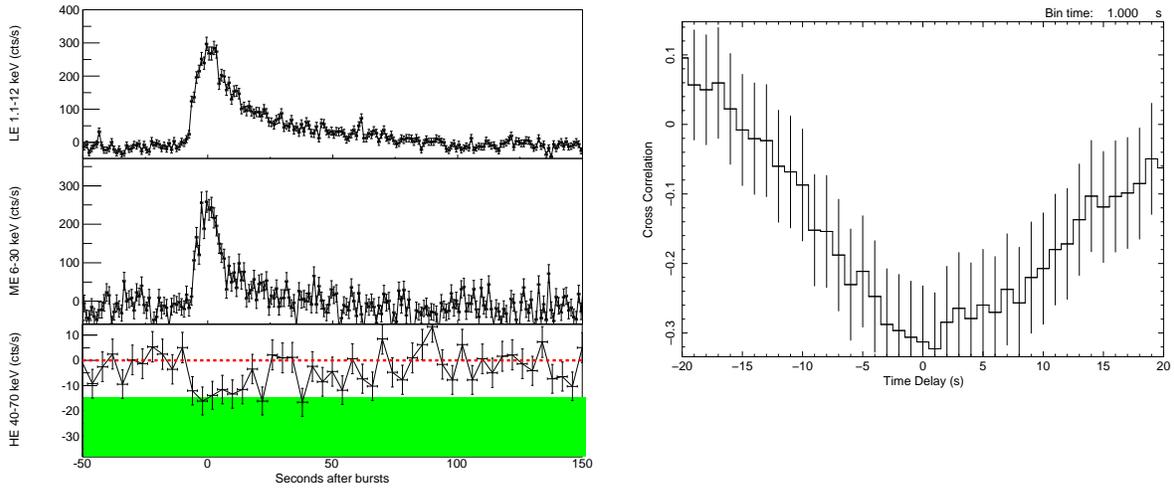}
      \includegraphics[origin=c,angle=270, scale=0.3]{cross.eps}
 \caption{The {\bf left panel} is  the LE, ME and HE lightcurves of the burst in 1.1-12 keV, 5-30 keV and 40-70 keV respectively, the time-bin for LE\&ME is 1s and HE is 4s, the green zone in bottom pad indicates the background level for HE detectors. The {\bf right} panel shows the cross-correlation between the left panels's LE and HE re-extracted lightcurves with time bin 1 s.}
\label{fig_lc}
\end{figure}

As shown in Fig. \ref{fig_lc}, the HE flux is mostly negative during the
burst and around zero elsewhere.
This deficit is 12.4$\pm$2.0 cts/s, and its significance is 6.2 $\sigma$, which is estimated by the ration of the deficit area  and  the sum of the error-bars with duration of 32 seconds.
The pre-burst emission of HE is $\sim$140 counts/s in 40-70 keV,
which include the background $\sim$125 counts/s and persistent emission  $\sim$15 counts/s.
From Fig. \ref{fig_lc}, the hard X-ray decrement reaches a maximum of 16.1$\pm$5.5 cts/s at the soft-X-ray burst peak.
Considering the two values above,  we conclude that  almost all of the persistent emission in 40-70 keV is crippled at the burst peak-time.

A cross-correlation analysis is attempted between the
LE lightcurves in 1.1-12 keV and re-extracted HE lighcurve in 40-70 keV with a bin
size of 1 s, as shown in Fig.\ref{fig_lc}.
The minimum of the cross-correlation value appears at 1.6 $\pm$1.2 s,
indicated that hard X-ray deficit delays the burst emission.
{\bf The value is derived from  Gaussian-fitting of Fig.\ref{fig_lc}.}


\subsection{Broad-band spectra of burst emission}

 The time-ordered burst spectra fitting results are as shown in Fig \ref{fig_fit}, the peak bolometric flux is $2.6\pm0.4\times10^{-8}~erg~cm^{-2}~s^{-1}$, which is roughly half of the Eddington luminosity  $6.0\pm0.6\times10^{-8}~erg~cm^{-2}~s^{-1}$ obtained from the PRE bursts of obsid P011465400301-20180213-01-01.
The reduced $\chi^{2}$  of the jointed spectra fitting are roughly at 1, with most of the degree of freedom more than 50.
The duration of the bursts is also estimated from the ration of the bolometric flux to the peak flux, $\sim$13.1 s.

\begin{figure}[b]
\begin{center}
 \includegraphics[origin=c, angle=0, scale=0.25]{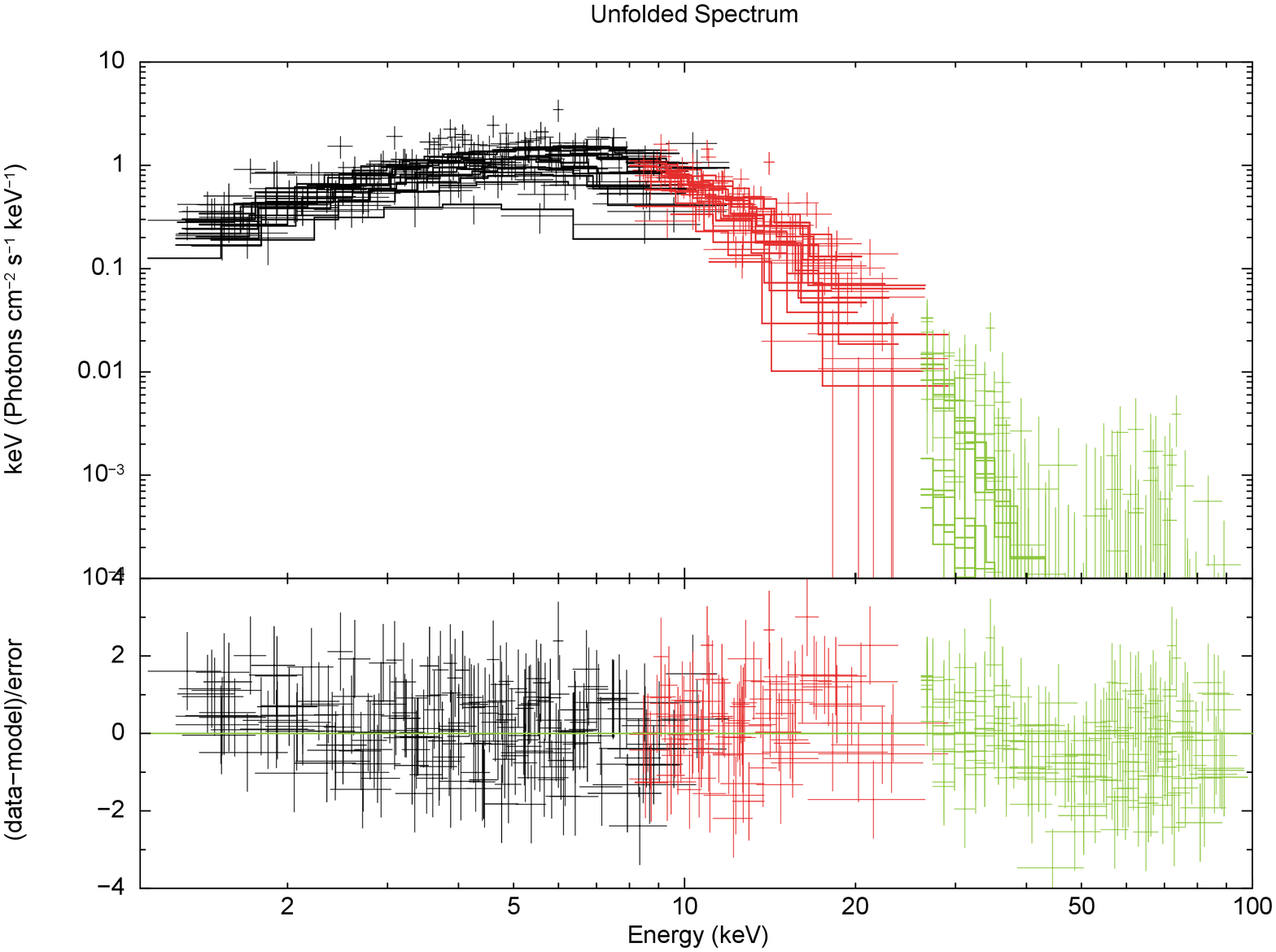}
  \includegraphics[origin=c, angle=0, scale=0.25]{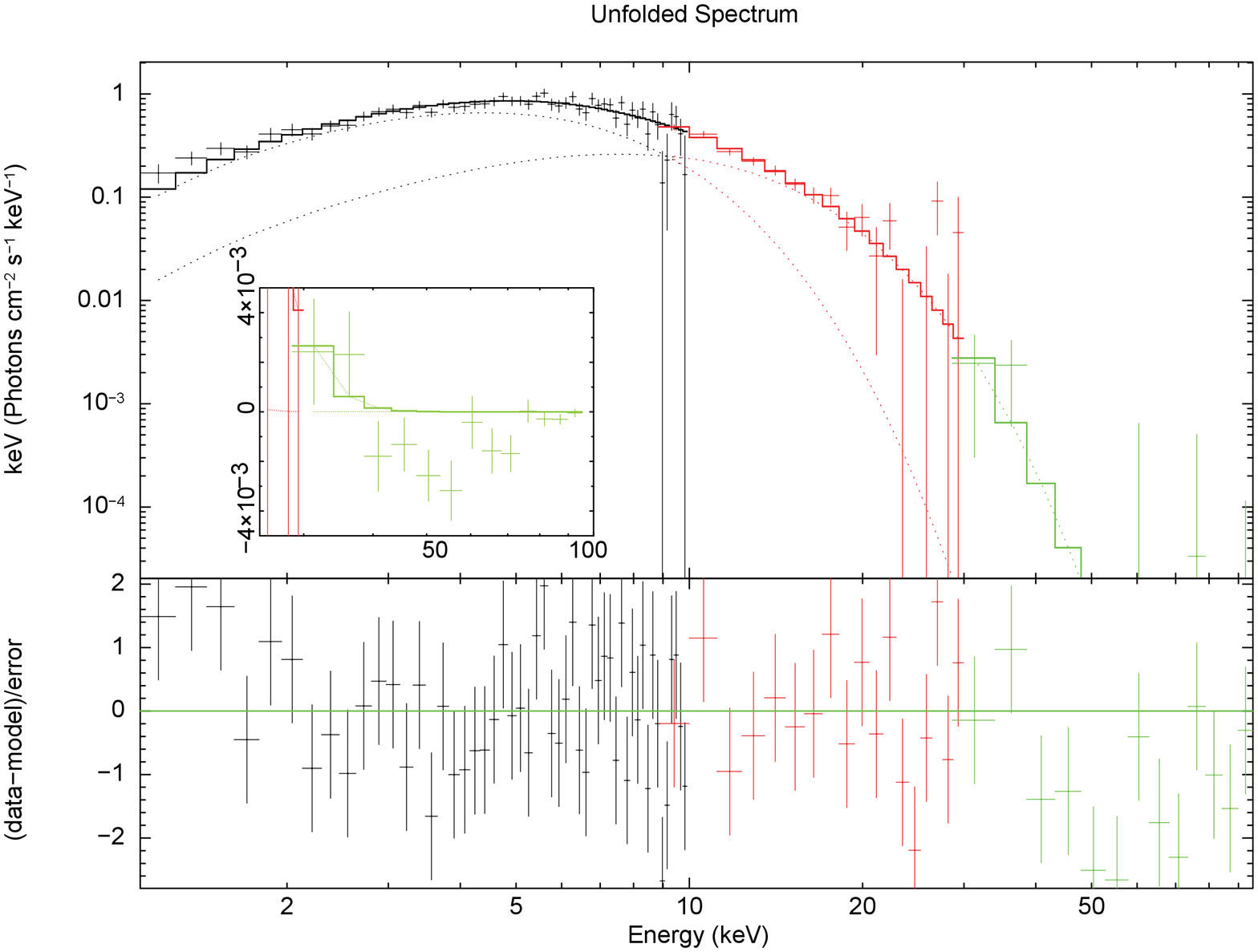}
 \caption{
 The {\bf left} panel is merged the 1s exposure burst spectra fit results to one graph for easy reading.
 The {\bf right} panel is 32s exposure time spectra during the burst with model  cons*wabs*(bbodryad+bbodyrad), the embedded panel shows  HE spectra in detail since the negative value can not been shown in main panel with logarithmic coordinate axis.
  The enegy range of the spectra of LE (black), ME (red) and HE (green) is 1.1-12 keV, 8-30 keV and 25-100 keV respectively.
 }
   \label{spec}
\end{center}
\end{figure}

In the process of  fitting 1-s exposure jointed spectra, by taking pre-burst emission as background,  while the residual of LE and ME are roughly around 0, a marginally deficit appears for HE.
For easy reading, all the time-ordered spectral fitting results in Fig \ref{fig_fit} are merged into one graph which is shown in Fig. \ref{spec},
a clear deficit is visible around 40-50 keV.
To investigate further the deficit, a jointed spectra with 32 seconds exposure time is extracted, and fitted
with two blackbody model (with absorption fixed).
The consideration of one more blackbody spectral component is to account for the temperature evolution in case of handing the entire burst.
The embedded HE spectra of Fig. \ref{spec} shows that the negative  values appear at 40-70 keV, the significance of this deficit in 40-70 keV from the spectra is 6.8 $\sigma$.
This value is estimated based on the different counts of the bursts and pre-burst emission from the spectra,  i.e., 4012 counts for the burst and 4465 counts for pre-burst emission at the same energy band.
From the above analysis, the significance of the deficit derived from the spectral and lightcurves are consistent each other.




\section{Discussion}

Observations of Insight/HXMT, provide us so far the best chance to study the  effects of type-I X-ray  burst on {\bf continuum} emission of NS XRB, thanks to the large detection area and wide bandwidth of the Insight-HXMT.  Here with the Insight-HXMT we find for the first time  the {\bf hard X-ray deficit/shortage} via  single short type-I burst showing up during {\bf flare} of 4U 1636-536.
 {\bf The} hard X-ray shortage is detected with significance 6.2  in 40-70 keV, and a cross-correlation analysis between the lightcurves of soft and hard X-ray band, shows that the corona shortage lag
the burst emission by 1.6$\pm$1.2 s. These results are consistent with those derived previously from stacking a large amount of bursts detected by XTE/PCA within a series of {\bf flares} of  4U 1636-536. Moreover,  the broad bandwidth of Insight-HXMT allows as well for the first time to infer the burst influence upon the continuum spectrum via performing the spectral fitting of the burst, which ends up with the finding that hard X-ray shortage appears at around 40 keV in the continuum spectrum.  These results suggest that the evolution of the corona along with the outburst{\bf /flare} of NS XRB may be traced via looking into a series of embedded type-I bursts by using Insight-HXMT.


Our previous work has revealed the deficit of the hard X-rays in six sources using RXTE/PCA data, by constructing a sample with tens bursts for each source (e.g., \citealp{chen2012,ji2013}, and reference therein).
Since all these reports are based on the RXTE/PCA observations, a suspicion may arise for the dead time concerning which may have influence upon the significance of  observing a hard X-ray shortage accompanied with the type-I burst.
Although later on INTEGRAL observations also confirm such kind of deficit in 4U~1728-34 \citep{Kajava2017}, through stacking a sample of 123 bursts in low/hard state,
but they reported  detection significances of  3.4 $\sigma$ in the 40-50 keV band and 1.8 $\sigma$ in the 50-80 keV band.
In this work, benefitting the broad  energy coverage of three different detectors,
this phenomenon is also detected for the first time with a single in single  burst of 4U~1636-536.


As were discussed in previous papers, the deficit in hard X-rays {\bf likely} indicates a cooling of the corona by the burst,
which provides an intense shower of the soft X-rays to  cool the hot corona via Compntonization .
The time lag between the burst at soft X-rays and the deficit at hard X-rays for the continuum  emission is considered as the timescale of the corona reheating/reformation.
So far all the time-lags of the deficits detected in the above sources are within several seconds,  indicating a similar mechanism for corona production during outburst/{\bf flare} of  NS XRBs.

\begin{figure}[t]
\centering
   \includegraphics[angle=0, scale=0.4]{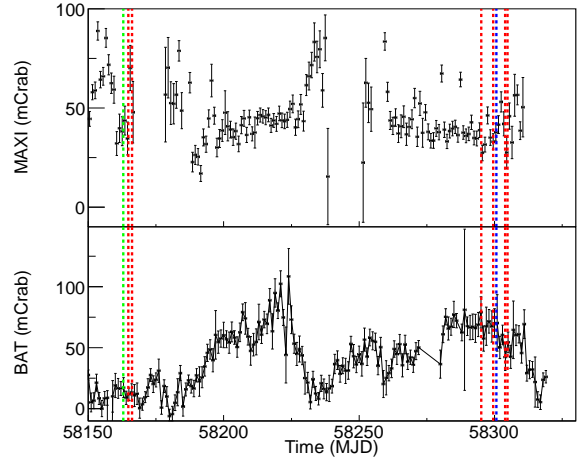}
 \caption{Daily lightcurves of 4U~1636-536 by MAXI and Swift/BAT during the outbursts in 2018, in 2-20 keV and 15-50 keV respectively. The burst is indicated by vertical lines, the PRE burst and the burst of this work are marked by green and blue respectively. }
\label{fig_outburst_lc}
\end{figure}

 As shown in Table \ref{table} and Fig. \ref{fig_outburst_lc}, apart form the burst with a hard X-ray shortage clearly seen, there are other three bursts detected during the soft state of the 4U 1636-536. Two of them are the short type-I bursts, and one is the PRE burst.  No hard X-ray shortages are detected for the continuum emission during these bursts, due to that the continuum hard X-ray emissions are too weak. As shown in the   Swift/BAT observations, the source stayed at roughly $\sim$ 15 mCrab level at above 15 keV  at time around these three bursts.  The spectral analysis of the these bursts, show no obvious significant residual detected in LE\&ME spectra fitting results.

 Although a hint of excess appears at below $\sim$ 2 keV for the burst showing up in the hard state of the {\bf flare}, it becomes invisible by setting the absorption parameter free during spectra fitting.
Usually, the persistent emission change are detected in RXTE/PCA and NICER, and the degree of enhancement is proportional to the burst flux.
The faint of the burst peak flux probably prevents us to detect the effects of the bursts on the disk emission.
Clear residuals below $\sim$ 2 keV are detected at the PRE burst of obsid P011465400301-20180213-01-01, which is detected with faint persistent emission, also indicates the state-depended of the persistent emission enhancement induced by the burst.
The PRE burst results will be reported by the upcoming paper.


The physical origin of QPOs at kHz is thought to be the dynamical timescale of the inner part of the accretion disk and hence provide another way to diagnostic the burst influence upon the accretion disk/corona.
For 4U~1636-536, QPO frequency changes during burst is  detected\citep{peille2014}.
One interpretation is that the inner part of the disk is puffed up by the burst, which suppresses the QPO generation \citep{Ballantyne2005}.
If the disk curls up to higher latitude, the deficit will become unremarkable because of cloaking the burst photons by the taller disk, since the disk is optically thick.
The other interpretation to the QPO suppression is the backward/regression of the inner disk, i.e., X-ray burst blow away the disk to behave bigger inner disk radius. However, the disk emission is detected to increase during burst by RXTE  \citep{int2013,Worpel2013,Ball2014,keek2014} and NICER \citep{Keek2018}, which indicates a smaller inner disk radius  induced by burst.
Unfortunately no QPO is detected from 4U 1636-536 by the Insight-HXMT,  probably due to its relatively small detection area at soft X-rays.
An increase of disk temperature maybe reconcile this contradiction, in other word, the burst heats the disk.

The deficit energy  covers a range of roughly 40-70 keV, extends to $\sim$70 keV, within which HE has the largest effective area.
Considering the effective area of HE decreases in higher energy and low counts rate of the pre-burst emission  for the continuum emission at hard X-rays, the deficit maybe both extends to lower/higher energies.
More observations on the low/hard state of 4U~1636-536 and target of opportunity (ToO) observations on brighter burster, such as Aql X-1, may provide us better chance to investigate thoroughly both the time and spectral  evolution of  corona along with outburst/{\bf flare} via investigating individual type-I burst with the unique capability of the Insight-HXMT.
{\bf The up-scattering photons of bursts by the continuum emission, should affect the shape/amplitude of blackbody spectra, joint observations on bursters by NICER or AstroSat and Insight/HXMT may give us opportunity to test the coronal cooling interpretation.}




\acknowledgments{This work is supported by the
National Key R\&D Program of China (2016YFA0400800)
and the National Natural Science Foundation of China under
grants 11473027, 11733009, U1838201 and U1838104.
}

\bibliographystyle{plainnat}


\begin{table}[ptbptbptb]
\begin{center}
\label{table}
\caption{The bursts obsid and peak time of 4U 1636-536  detected by Insight/HXMT. }
\begin{tabular}{cccccccccccccccccc}
\\\hline
 No  & obsid	& Time (MJD) \\\hline
1${\mathrm{*}}$	 &	P011465400301-20180213-01-01	 &	58162.871091	\\
2	 &	P011465400401-20180215-01-01	 &	58164.733723	\\
3	 &	P011465400501-20180217-01-01	 &	58166.117179	\\
4	 &	P011465402301-20180626-01-01	 &	58295.087021	\\
5	 &	P011465402701-20180630-01-01	 &	58299.489308	\\
6	 &	P011465402801-20180701-01-01	 &	58300.717896	\\
7	 &	P011465403101-20180704-01-01	 &	58303.959479	\\
8	 &	P011465403201-20180705-01-01	 &	58304.808113	\\
\hline
\end{tabular}
\end{center}
\begin{list}{}{}
\item[${\mathrm{*}}$]{Burst shows photosphere radius expansion.}
\end{list}
\end{table}


\end{document}